\documentclass[aps,prl,twocolumn,superscriptaddress,showpacs]{revtex4}
\usepackage{graphicx}
\usepackage{amsmath}
\usepackage{natbib}
\begin{document}
\title{Signatures of homoclinic motion in quantum chaos}
\author{D. A. Wisniacki}
\affiliation{Departamento de Qu\'{\i}mica C--IX,
 Universidad Aut\'onoma de Madrid,
 Cantoblanco, 28049--Madrid, Spain.}
\affiliation{Departamento de F\'\i sica ``J. J. Giambiagi'', FCEN,
UBA,
 1428 Buenos Aires, Argentina.}
\author{E. Vergini}
\affiliation{Departamento de Qu\'{\i}mica C--IX,
 Universidad Aut\'onoma de Madrid,
 Cantoblanco, 28049--Madrid, Spain.}
\affiliation{Departamento de F\'{\i}sica,
 Comisi\'on Nacional de Energ\'{\i}a At\'omica.
 Av.\ del Libertador 8250, 1429 Buenos Aires, Argentina.}
\author{R. M. Benito}
\affiliation{Departamento de F\'{\i}sica,
 E.T.S.I.\ Agr\'onomos, Universidad Polit\'ecnica de Madrid,
 28040 Madrid, Spain.}
\author{F. Borondo}
\email[E--mail address: ]{f.borondo@uam.es}
\affiliation{Departamento de Qu\'{\i}mica C--IX,
 Universidad Aut\'onoma de Madrid,
 Cantoblanco, 28049--Madrid, Spain.}
\date{\today}
\begin{abstract}
Homoclinic motion plays a key role in the organization of classical
chaos in Hamiltonian systems.
In this Letter, we show that it also imprints a clear signature in
the corresponding quantum spectra.
By numerically studying the fluctuations of the widths of
wavefunctions localized along periodic orbits
we reveal the existence of an oscillatory behavior,
that is explained solely in terms of the primary homoclinic motion.
Furthermore, our results indicate that it survives the semiclassical
limit.
\end{abstract}
\maketitle
%

Periodic orbits (POs) are invariant classical objects of outmost
relevance for the understanding of chaotic dynamical systems.
The pioneering work of Poincar\'e demonstrated the importance of these
retracing orbits in organizing the complexity of the corresponding
classical motion.
Semiclassically, Gutztwiller \cite{Gutzwiller} developed his
celebrated trace formula, able to quantize chaotic systems,
solely in terms of POs information.
Quantum mechanically, striking manifestations of POs have been
reported in the literature.
One of the most important are ``scars'' \cite{Heller1},
an enhanced localization of quantum density along
unstable short POs in certain individual eigenfunctions.
This phenomenon was first noticed in the Bunimovich stadium
billiard \cite{McDonald}, and subsequently studied systematically
by Heller, who constructed a theory of scarring based on wave packet
propagation \cite{Heller2}.
Another important contribution to scar theory is due to Bogomolny
\cite{Bogomolny}, who derived an explicit expression for the PO
contributions to the smoothed quantum probability
density over small ranges of space and energy
(i.e.\ average over a large number of eigenfunctions).
A corresponding theory for Wigner functions was developed by Berry
\cite{Berry}.
Recently, there has been a flurry of activity focusing on the
influence of bifurcations (mixed systems) on scarring \cite{Keating}.

The existence of a scar implies a clear regularity in the corresponding
quantum spectrum, related to the period of the PO.
In time domain, the dynamics of a packet running along a PO induce
recurrences in the autocorrelation function, that when Fourier
transformed define an envelope in the spectrum, giving rise to peaks
of width proportional to the Lyapunov exponent, $\lambda$,
at energies given by a Bohr--Sommerfeld (BS) quantization condition
\cite{Gutzwiller,Heller3}.

In this Letter, we demonstrate the existence of an additional
superimposed spectral regularity also related to scarring.
It is originated by the associated homoclinic motion and is given by
the area enclosed by the stable and unstable manifolds
up to the first crossing.
To unveil this regularity we consider the fluctuations of the spectral
widths corresponding to localized wavefunctions along unstable POs.
Our data demonstrate that these fluctuations have a surprisingly
simple oscillatory behavior essentially governed by the
quantization of only the primary homoclinic dynamics.
We provide an explanation of this result in terms of the coherence
of this classical motion \cite{Ozorio}, which constitute the natural
global extension of the local hyperbolic structure around the PO.
Furthermore, our numerical results indicate that the observed oscillatory
behavior do not vanish as $\hbar \rightarrow 0$.

It should be remarked that,
contrary to the previously described results on scar theory,
our study implies dynamics beyond the Ehrenfest time ($\sim|\ln \hbar|$),
when the PO manifolds start to cross and the homoclinic tangle develops,
giving rise to subtle quantum interference effects.
Other interesting papers, also considering these longer times, are due
to Tomsovic and Heller \cite{Tomsovic}, who showed how to construct a
valid semiclassical approximation to wavefunctions past this limit;
this came as a surprise since the underlying chaos has had time to
develop much finer structure than a quantum cell.
Similarly to what happens in energy domain with Gutzwiller trace formula,
this theory unfortunately requires the use of a number of homoclinic
excursions increasing exponentially with time.
Our result indicates that there exists some properties of these long term
dynamics that can be understood in terms of a small number of classical
invariants.

The ideal tool to carry out our investigation are non--stationary wave
functions highly localized on POs.
The overlap with the corresponding eigenstates provides information on
how these structures appear in the spectra, and how they are embedded
in the quantum mechanics of the system.
Such functions, associated to a PO, can be obtained by using the
(dynamical) information up to the Ehrenfest time,
thus including the hyperbolic structure formed by the corresponding
unstable and stable manifolds.
In this way, the associated wavefunctions spread along these dynamically
relevant regions of phase space (see illustration below in Fig.~\ref{fig:1}).
This can be done in different ways \cite{Polavieja,Vergini2};
in particular, in Ref.\ \onlinecite{Vergini2} a definition of scar functions,
based on transversally excited resonances along POs at given BS
quantized energies with minimum dispersion, is provided.
The width of these functions in the energy spectrum is given by,
\begin{equation}
   \sigma_{\rm sc} = \frac{\pi\hbar\lambda}{|\ln \hbar|}+
     {\cal O} \left(\frac{\hbar}{|\ln \hbar|^2} \right),
 \label{eq:sigmasc}
\end{equation}
where the narrowing factor $|\ln \hbar|^{-1}$ comes from the inclusion
of the hyperbolic structure \cite{Vergini2}.
In this respect, it should be noticed that although this approach
incorporates the motion along the manifolds up to the Ehrenfest time,
interference effects due to the intersections of the manifolds are
not included.

The model used in our calculations is the fully chaotic system
consisting of a particle of mass $M$ confined in a desymmetrized
Bunimovitch stadium billiard, defined by the radius of the
circular part, $r=1$, and the enclosed area, $1+\pi/4$. To
calculate the corresponding eigenstates, Dirichlet conditions on
the stadium boundary and Neumman conditions on the symmetry axes
are imposed.
We will focus our attention on the dynamics influenced by the horizontal PO.
The corresponding scar wavefunctions, $|\gamma\rangle$, are calculated
using the method of Ref.~\cite{Vergini2}.

In Fig.~\ref{fig:1} we show, as an example, the results for the scar
state with wavelength $k_{\rm BS}=211.665$. This value was obtained
from the BS rule, $k_{\rm BS}=(2\pi/L_H)(n+\nu_H/4)$, where $n=134$
is the excitation number along the orbit, $L_H=4$ its length,
and $\nu_H=3$ the corresponding Maslov index.
\begin{figure}[t]
 \includegraphics[width=7.5cm]{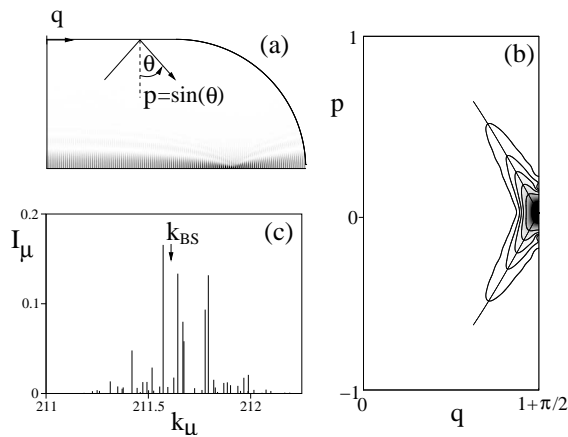}
 \caption{Configuration space (a), phase space (b), and
   wavelength spectrum weights (c) representations of a scar wavefunction
   along the horizontal periodic orbit of a desymmetrized stadium
   billiard corresponding to BS wavelength $k_{\rm BS}=211.665$.
   The unstable and stable manifolds of the orbit are also plotted
   in panel (b).
   In panel (a) the Birkhoff coordinates on the boundary of the
   billiard  are shown.}
 \label{fig:1}
\end{figure}
In it, all features discussed above are observed. Namely, the
Husimi based quantum surface of section \cite{Wisniacki1} spreads
along the manifolds structure associated to the horizontal PO, and
the state appears in the spectrum distributed among the different
eigenstates, $|\mu\rangle$, within a given range around the
corresponding BS quantized wavelength, with intensity weights
$I_{\mu} \equiv |\langle \mu|\gamma\rangle|^2$.

\begin{figure}[t]
 \includegraphics[width=6.5cm]{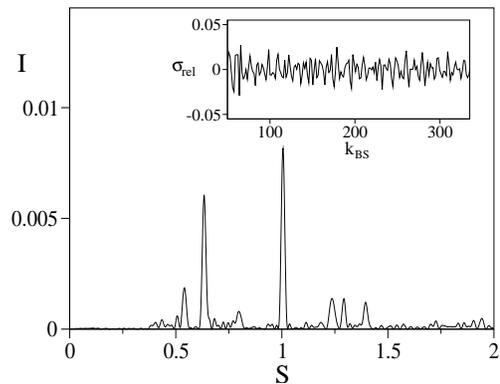}
 \caption{Relative width for the scar states along the horizontal
  periodic orbit as a function of the Bohr--Sommerfeld quantized
  wavelength (inset), and its Fourier transform.}
 \label{fig:2}
\end{figure}

The energy width of the associated envelope can then be defined as
(we take $M=\hbar^2/2$ in such a way that $k^2$ is the energy of
the particle)
\begin{equation}
   \sigma = \sqrt{\sum_\mu \; |\langle \mu|\gamma\rangle|^2 \;
            (k_\mu^2 - k_{\rm BS}^2)^2}.
 \label{eq:sigma}
\end{equation}
When calculated, this quantity is a decreasing oscillatory function
of $k_{\rm BS}$, that can be more adequately studied by defining its
relative variation with respect to the corresponding semiclassical value,
\begin{equation}
  \sigma_{\rm rel} \equiv \frac{\sigma-\sigma_{\rm sc}}{\sigma_{\rm sc}}.
 \label{eq:sigmarel}
\end{equation}
The corresponding results are shown in the inset to Fig.~\ref{fig:2}.
%
As can be observed, it exhibit an oscillatory behavior, with an
amplitude representing only a small fraction of $\sigma_{\rm sc}$.
When this signal is Fourier analyzed (see main body of Fig.~\ref{fig:2})
it is seen to have only two relevant components, corresponding
to the peaks appearing at values of the action $S=0.633$ and 1.007,
respectively.
(Notice that, in our case, $S$ has units of length since the total
linear momentum of the particle has been set equal to one).
The BS quantization of the wavelength implies that
the Fourier transform of $\sigma_{\rm rel}$ is a periodic function
with period $L_H$. In this respect, corresponding peaks appear at
$S=-3.367$ and $-2.993$.
%
\begin{figure*}
 \includegraphics[width=12.0cm]{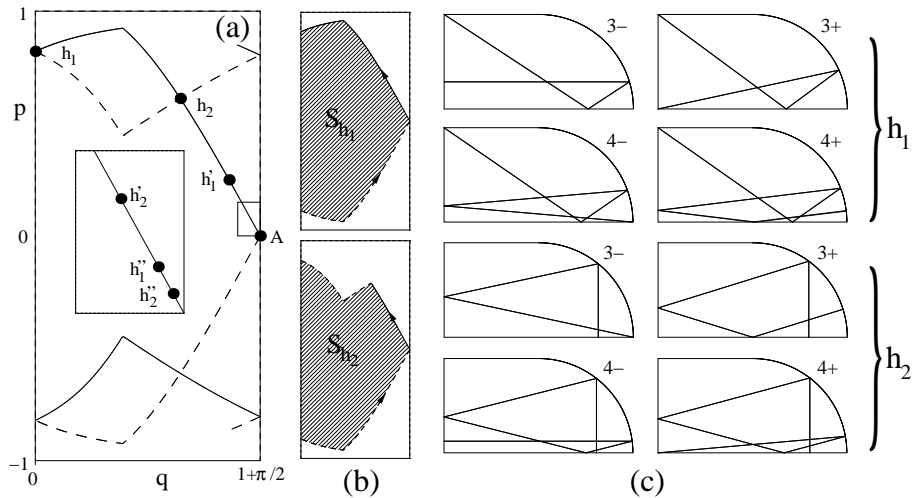}
 \caption{(a) Phase space portrait in Birkhoff coordinates relevant
  to our calculations.
  Label A indicates the position of the fixed point corresponding to
  the horizontal periodic orbit.
  The associated unstable and stable manifolds are represented
  in full and dashed line, respectively.
  They cross, with different topology, at the primary homoclinic
  points ${\rm h}_1$ and ${\rm h}_2$, which map into
  ${\rm h}'_1$, ${\rm h}''_1$, $\ldots$,
  and ${\rm h}'_2$, ${\rm h}''_2$, $\ldots$. \protect \\
  (b) Primary homoclinic areas $S_{{\rm h}_1}$ and $S_{{\rm h}_2}$
  associated to the horizontal orbit. \protect \\
  (c) Satellite periodic orbits converging to the homoclinic points
  ${\rm h}_1$ and ${\rm h}_2$. See text for details.}
 \label{fig:3}
\end{figure*}

To interpret this result, we consider only the primary homoclinic
motion corresponding to the horizontal PO. This follows the
analysis of Ozorio de Almeida \cite{Ozorio}, who studied the
quantization of homoclinic orbits, that were thought of as
defining an invariant torus, approached by series of satellite
unstable POs. There are two topologically distinct types of
primary homoclinic points, which appear as the first and second
intersections of the associated manifolds. The situation is shown
in Fig.~\ref{fig:3} (a), where a picture of the relevant phase
space, in Birkhoff coordinates, is presented. In it, we have
plotted the fixed point corresponding to the horizontal PO
(labelled A), and the emanating unstable (full line) and stable
(dashed line) manifolds. These manifolds first cross (with
different topology) at points ${\rm h}_1$ and h$_2$, and
afterwards at the sequences ${\rm h}'_1$, ${\rm h}''_1$, $\ldots$,
and ${\rm h}'_2$, ${\rm h}''_2$, $\ldots$ (and their reflections
with respect to $p=0$), as fixed point A is approached. These
infinite sequences of points, which are the dynamical images of
${\rm h}_1$ and ${\rm h}_2$, constitute the primary homoclinic
orbits, that define two relevant areas in phase space, denoted
hereafter as $S_{{\rm h}_1}$ and $S_{{\rm h}_2}$ [see shaded
regions in Fig.~\ref{fig:3} (b)], important for the discussions
presented below. Furthermore, these primary homoclinic orbits
accumulate infinite sets of satellite POs with increasingly longer
periods, as discussed in Ref.~\cite{Ozo2}. The first members of
these two families, converging to the primary homoclinic orbits
are presented in Fig.~\ref{fig:3} (c). Accordingly to our
notation, based on the number of times that the POs intersect our
surface of section, orbits 3-- and 3+ consist of fixed points near
${\rm h}_1$, ${\rm h}'_1$ (or ${\rm h}_2$, ${\rm h}'_2$), and
their reflections with respect to $p=0$; orbits 4-- and 4+
consists of fixed points near ${\rm h}_1$, ${\rm h}'_1$, ${\rm
h}''_1$ (or ${\rm h}_2$, ${\rm h}'_2$, ${\rm h}''_2$), and their
reflections; and so on for the longer POs in the family. Notice
that the reflection with respect to $p=0$ of points over $q=0$ or
$q=1+\pi/2$ gives the same point. Moreover, these POs can be
grouped in pairs, in such a way that 3-- and 3+ represent the
first approximation to the primary homoclinic motion (the --
indicates the PO with shortest length of the pair), 4-- and 4+ the
second, etc. Actually, it is apparent the increasing relation of
these orbits to the horizontal one, since as we progress in the
sequence new segments closer and closer to the horizontal axis are
incorporated into the trajectory.

Now, the question arises about under which conditions the satellite
families of POs reinforce the contribution of the central one.
As seen before, the horizontal PO satisfy the BS quantization condition
\begin{equation}
    k L_H - \frac{\pi}{2} \nu_H = 2 \pi n_H,
 \label{eq:BSH}
\end{equation}
being $n_H$ an integer.
In the same way, the BS condition for orbits crossing $m$ times the
surface of section reads
\begin{equation}
    k L_{m} - \frac{\pi}{2} \nu_{m} = 2 \pi n_{m}.
 \label{eq:BS12}
\end{equation}
Subtracting $m$ times Eq.~(\ref{eq:BSH}) to (\ref{eq:BS12}),
a new quantization condition emerges
\begin{equation}
   k(L_{m}-m L_H)-\frac{\pi}{2}(\nu_{m}-m \nu_H)=2\pi(n_{m}-m n_H).
 \label{eq:BScombi}
\end{equation}

The relevant point here is that $L_m-m L_H$ converges exponentially
to $S_{{\rm h}_1}=-3.368\;390\;45$ and $S_{{\rm h}_2}=-2.991\;142\;22$
for the two families ${\rm h}_1$ and ${\rm h}_2$, respectively,
as shown in Table \ref{table:I}.
Moreover, $\nu_m-m\nu_H$ is independent of $m$, and it corresponds to
$\nu_{{\rm h}_1}=-1$ and $\nu_{{\rm h}_2}=0$.
In this way, we are in the position to assess that all satellite orbits
of a given family contribute coherently when two BS quantization rules
are simultaneously fulfilled:
(a) the quantization of the central (horizontal) orbit
    [Eq. (\ref{eq:BSH})], and
(b) the so--called quantization of the homoclinic torus according to
    Ref.~\cite{Ozorio},
\begin{equation}
    k S_{{\rm h}_i} - \frac{\pi}{2} \nu_{{\rm h}_i} = 2 \pi n,
       \qquad i=1,2.
 \label{eq:BSSH1}
\end{equation}
%
\begin{table}
 \begin{center}
 \begin{tabular}{cccc}
  \hline  \hline
   $m$ & \multicolumn{3}{c}{$\overline{L}_m-mL_H$} \\
  \hline
       & Family ${\rm h}_1$ & & Family ${\rm h}_2$ \\
  \hline
    3  & -3.367 727 48 & & -2.990 915 39 \\
    4  & -3.368 367 57 & & -2.991 131 87 \\
    5  & -3.368 389 68 & & -2.991 141 81 \\
    6  & -3.368 390 43 & & -2.991 142 21 \\
    7  & -3.368 390 45 & & -2.991 142 22 \\
  \hline  \hline
 \end{tabular}
 \end{center}
 \caption{Exponential convergence of $\overline{L}_m-m L_H$ as
   a function of the number of times  the POs
   intersect our surface of section
   for families ${\rm h}_1$ and ${\rm h}_2$ of
   satellite orbits corresponding to the primary homoclinic motion.
   To improve the convergence we have averaged the length of orbits
   $m$+ and $m$-- for the same family [see Fig.~\ref{fig:3} (c)].}
 \label{table:I}
\end{table}
It should be observed that the values of $S_{{\rm h}_1}$ and
$S_{{\rm h}_2}$ agree extremely well with those obtained numerically
for the position of the two main peaks in the spectrum of
Fig.~\ref{fig:2}.
On the other hand, these homoclinic areas correspond in our chosen
Poincar\'e surface of section to the areas indicated in
Fig.~\ref{fig:3} (b).  The arrows in Fig.~\ref{fig:3} (b)
indicate the anti-clockwise direction of the homoclinic motion,
and for this reason homoclinic areas result negative.

This notorious influence of the homoclinic motion in the spectra,
and thus in the quantum mechanics of our system, is remarkable,
especially, when one takes into account that the homoclinic dynamics
is the origin of the chaotic behavior in Hamiltonian systems.
In order to understand this behavior we note that the motion
has two components.
The first one, close to the primary homoclinic torus,
is such that the corresponding trajectories interfere coherently,
as it has been shown.
The second component, associated to the non-primary homoclinic crosses,
gives rise to more complicated self--intersecting tori.
These excursions are much longer and the corresponding interference
is much less coherent;
accordingly, its influence in the spectra is very diluted.
In this respect, it is more efficient to consider these
long excursions, by the interaction  with other short POs
(heteroclinic connections).
This point has been discussed elsewhere \cite{hetero}.

\begin{figure}[t]
 \includegraphics[width=6.5cm]{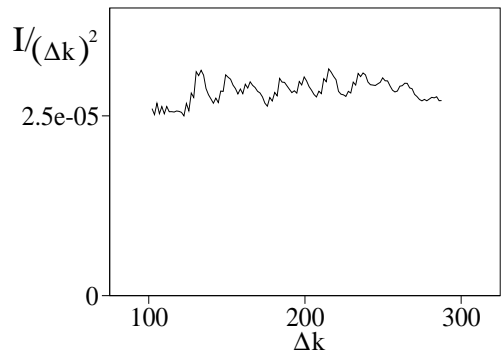}
 \caption{Intensity of the peak at $S_{{\rm h}_1}$ divided by
   $(\Delta k)^2$ as a function of $\Delta k$}.
 \label{fig:4}
\end{figure}

Finally, let us consider if the influence of the homoclinic motion
on $\sigma_{\rm rel}$ survives the semiclassical limit
$\hbar\rightarrow 0$ ($k=\frac{1}{\hbar} \rightarrow \infty$ in
our units). For this purpose we have performed the following
calculation. Starting from the scaled fluctuation curve
$\sigma_{\rm rel}$, presented in the inset to Fig.~\ref{fig:2}, we
compute the Fourier transform of the function for increasingly
larger intervals of $k_{\rm BS}$, $(k_0,k_1)$, keeping fixed the
value of the lower limit, $k_0$. The results for the peak at
$S_{{\rm h}_1}$ are shown in Fig.~\ref{fig:4}, where it is seen
that the corresponding Fourier intensity divided by $(\Delta k)^2$
is an approximately constant function of the length of the
interval, $\Delta k=k_1-k_0$. A similar behavior is obtained for
the peak at $S_{{\rm h}_2}$. This result implies that $\sigma_{\rm
rel} (k_{\rm BS})$ is essentially an oscillatory function with
frequencies $S_{{\rm h}_1}$ and $S_{{\rm h}_2}$ and constant
coefficients.
%

In conclusion, we have gone one step further the usual
understanding of the quantum--classical correspondence, by showing
how homoclinic areas imprint a clear signature in the spectra of
classically chaotic systems. In particular, we have found that the
fluctuations of the relative widths corresponding to wavefunctions
(with hyperbolic structure) highly localized in the vicinity of a
PO oscillate in a very simple way, that is clearly controlled by
the quantization of the associated primary homoclinic motions.
This indicates the importance of classical invariants, other than
those included in Gutzwiller theory concerning individual POs, for
the systematic description of quantum information.
%
\begin{acknowledgments}

This work was supported by MCyT and MCED (Spain) under contracts
BQU2003--8212, SAB2000--340, and SAB2002--22.
\end{acknowledgments}
%

%

\begin{thebibliography}
\eprint{}

\bibitem{Gutzwiller} M. C. Gutzwiller,
  {\it Chaos in Classical and Quantum Mechanics}
  (Springer--Verlag, New York, 1990).

\bibitem{Heller1} E. J. Heller, Phys. Rev. Lett. {\bf 53}, 1515 (1984).

\bibitem{McDonald} S. W.  McDonald,
 {\it LBL Report} 14837 (1983).

\bibitem{Heller2} L. Kaplan and E. J. Heller,
  Ann. Phys. {\bf 264}, 171 (1998).

\bibitem{Bogomolny} E. B. Bogomolny,  Physica D {\bf 31}, 169 (1988).

\bibitem{Berry} M. V. Berry, Proc. R. Soc. Lon. A {\bf 243}, 219 (1989).

\bibitem{Keating} J. P. Keating and S. D. Prado,
  Proc. R. Soc. Lon. A {\bf 457}, 1855 (2001).

\bibitem{Heller3} E. J. Heller, in
  {\it Chaos and Quantum Physics},
  edited by M. J. Giannoni, A. Voros, and J. Zinn--Justin
  (Elsevier, Amsterdam, 1991).

\bibitem{Ozorio} A. M. Ozorio de Almeida,
  Nonlinearity {\bf 2}, 519 (1989).

\bibitem{Tomsovic} S. Tomsovic and E. J. Heller,
  Phys. Rev. Lett. {\bf 70}, 1405 (1993);
  Phys. Rev. E. {\bf 47}, 282 (1993).

\bibitem{Polavieja} G. G. de Polavieja, F. Borondo, and R. M. Benito,
  Phys. Rev. Lett. {\bf 73}, 1613 (1994).

\bibitem{Vergini2} E. G. Vergini and G. Carlo,
  J. Phys. A {\bf 34}, 4525 (2001).


\bibitem{Wisniacki1}
  D. A. Wisniacki, F. Borondo, E. Vergini, and R. M. Benito,
  Phys. Rev. E {\bf 63}, 66220 (2001).

\bibitem{Ozo2} G. L. Da Silva Ritter, A. M. Ozorio de Almeida,
  and R. Douady, Physica D {\bf 29}, 181 (1987).

\bibitem{hetero}
  D. A. Wisniacki, F. Borondo, E. Vergini, and R. M. Benito,
    Phys. Rev. E \textbf{70}, 035202(R) (2004).
%
\end{thebibliography}
\end{document}